# PRINCIPLES OF THE THEORY OF VORTEX GRAVITATION

S. A. Orlov

Petrozavodsk State University

The developed theory proves that a universal vortex motion, along with the pressure variation in a space continuum called 'ether', is actually the source of the universal gravitation and creation of celestial bodies and their motion in the Universe.

Calculations of the gravitation forces are carried out on the basis of the laws of mechanics of continua and (or) aerodynamics with the use of the Navier-Stokes equations.

As a result of the solution, an algebraic formula for the gravitation forces is obtained. The reliability of this formula is supported by its conformity to the astronomical data, and also to the Newton's empirical formula for universal gravitation.

The main advantage of the proposed theory is that, on the basis of it, it is possible to explain all the phenomena and laws observed in the cosmic space, viz.: motion of celestial bodies; mutual closing and moving away of galaxies; an origin of celestial bodies, black holes, and the Universe as a whole; the nature of gravity; the speed of gravitation; the strengths of magnetic fields of celestial bodies, etc.

## 1. Principles of the Theory

The proposed principle of the universal attraction forces is worked out on the following basis: Cosmic space is filled with a cosmic matter (ether) which forms an infinite spatial system of torsion vortexes. The physical characteristics of the ether are [1]:

**density** – $8{,}85 \cdot 10^{-12}$ kg/m$^3$,

**pressure** – $2 \cdot 10^{32}$ Pa,

**temperature** – $7 \cdot 10^{-51}$ K.

## 2. Vortex Gravitation

In the present work, we do not consider the origin of the cosmic vortexes. However, one can assume that the pressure decrease and the ether vorticity in space are caused by some thermodynamic processes, particularly, by flashes and explosions of the space matter.

Ether vortexes are characterized by the power and volume, which may be of any magnitude. Each vortex originates in rotation orbits of another, larger vortex.

**A funnel-shaped decrease of the pressure in the torsion creates the gravitation force,** which ensures the accumulation of the space matter in the central region of the torsion and hence – the creation of any celestial body.

Vortex gravitation, along with centrifugal forces, ensures the regular rotary motion of all the celestial bodies and systems around either the axis or another body; it also determines the gravity on the surfaces of planet, satellites, and stars, and, consequently, it determines the structure of the Universe.

**The action of the gravitation forces obeys the laws of aerodynamics.**

### 2.1. Model of the Origin of the Universal Gravitation Force

In this section, a model of appearance of the gravitation attraction force is considered from the viewpoint of aerodynamics. Namely, the two-dimensional model (Fig.1) is considered on the basis of the following initial postulates. These postulates will be expanded and defined more exactly below.

1. There exists an ether vortex around any physical object.

2. The ether motion in the vortex has laminar nature and obeys the laws of hydro- or aerodynamics; the ether viscosity is low.

3. The pressure gradient, arising during the vortex motion of the ether gas, is the reason for attractive force from body 1 to body 2 (see Fig.1).

4. The direction of the force $F_п$ does not depend on the direction of the ether angular velocity, which is necessary for the attractive force between the bodies, irrespective of their relative position. This implies the absence of the Magnus force – the force of interaction between two vortexes which appears in the classical aerodynamics. Such assumption can take place in the case of a weak interaction between the two ether flows, as if they would move one through another, not affecting mutual motion.

5. The appearing attraction force must describe the experimentally obtained law of gravity:

$$F_п = G \cdot \frac{m_1 \cdot m_2}{r^2} \qquad (1)$$

where $m_1$, $m_2$ are the masses of bodies 1 and 2, respectively, $G=6.672 \cdot 10^{-11}$ N·m²/kg² – the gravitational constant, and r – the distance between the bodies.

Next we consider the appearance of the attraction force in more detail and derive a formula describing it. As was said above, a pressure gradient arises as the result of the vortex motion. Let's find the radial distribution of the pressure and the ether velocity. For this purpose, we write the Navier-Stokes equation for the motion of a viscous liquid (gas).

$$c\left[\frac{\partial}{\partial t} + \vec{v}\cdot \text{grad}\right]\vec{v} = \vec{F} - \text{grad } P + з Д\vec{v} \qquad (2)$$

where ρ is the ether density, $\vec{v}$ and P are, respectively, its velocity and pressure, and η - the ether viscosity. In cylindrical coordinates, taking into account the radial symmetry $v_r=v_z=0$, $v_φ=v(r)$, $P=P(r)$, the equation can be written as the system:

$$\begin{cases} -\dfrac{v(r)^2}{r} = -\dfrac{1}{c}\dfrac{dP}{dr} \\ з \cdot \left(\dfrac{\partial^2 v(r)}{\partial r^2} + \dfrac{\partial v(r)}{r\partial r} - \dfrac{v(r)}{r^2}\right) = 0 \end{cases} \qquad (3)$$

In case of a compressible substance (ether), there will be a function $c = f(P)$ (instead of ρ).

From the first equation of system (3), one can find P(r) provided that the dependence v(r) is known. The latter, in turn, should be found from the second equation of that same system (one of the solution of which is the function v(r) ~ 1/r). At zero viscosity, the system permits any dependence v(r) [2].

The force affecting the body can be estimated from the formula

$$\vec{F}_п = - V \bullet \text{grad } P(r) \qquad (4)$$

where V is the volume of body 2.

In cylindrical coordinates the modulus of $\vec{F}_n$ is

$$F_n = V \cdot \frac{\partial P}{\partial r} \quad (5)$$

Then, comparing equations (3) and (5), for the incompressible ether (ρ=const) we find that

$$F_n = V \cdot c \cdot \frac{v(r)^2}{r} \quad (6)$$

In order that $F_n(r)$ would correspond to the law of gravity (see Postulate 5), $v(r)$ must obey the dependence $v(r) \sim \frac{1}{\sqrt{r}}$, and not the $v(r) \sim \frac{1}{r}$.

Taking into account the edge condition $v(r_1) = w_1 \cdot r_1$,

$$v(r) = \frac{w_1 \cdot r_1^{\frac{3}{2}}}{\sqrt{r}} . \quad (7)$$

Thus,

$$F_n = V \cdot c \cdot \frac{w_1^2 \cdot r_1^3}{r^2} \quad (8)$$

Here we make one more supposition (№ 6) – Ether penetrates through all the space, including the physical bodies. The volume V in formula (8) is an effective volume, i.e. the volume of elementary particles, which the body is composed of. All the bodies are composed of electrons, protons, and neutrons. The radius of an electron is much smaller that that of a proton and neutron. The radii of the latter are approximately equal to each other, $r_n \sim 1.2 \cdot 10^{-15}$ m. The same is true as to the masses: $m_n \sim 1.67 \cdot 10^{-27}$ kg ($r_n$ and $m_n$ are the radius and the mass of a nucleon). Therefore, the volume in formula (8) is:

$$V = \frac{m_2}{m_n} \cdot \frac{4p}{3} \cdot r_n^3 \quad (9)$$

Taking formula (9) into account, Eq.(8) can be rewritten as

$$F_n = \frac{4 \cdot p \cdot r_n^3 \cdot c}{3 \cdot m_n} \cdot \frac{w_1^2 \cdot r_1^3 \cdot m_2}{r^2} \quad (10)$$

Supposing further (supposition № 7) that

$$w_1^2 \cdot r_1^3 = A \cdot m_1 \qquad (11)$$

where A is a constant, Eq.(10) takes the form

$$F_п = \frac{4 \cdot p \cdot r_n^3 \cdot c}{3 \cdot m_n} \cdot A \cdot \frac{m_1 \cdot m_2}{r^2} \qquad (12)$$

Comparing equations (12) and (1), one can find that A=1.739·10¹⁸ m³/s²·kg. For the calculations, the data concerning the parameters of free ether, given in Section 1, were used.

The supposition № 7 is reasonable, since $w_1$ and $r_1$ are the parameters of body 1. If we divide both the left- and right-hand side of Eq.(11) by $r_1^3$, we get that the square of the ether angular velocity on the surface is proportional to the body's density. Let's find, e.g., the angular ether velocity on the surface of the Sun:

$$w_1 = \sqrt{A \cdot \frac{m_1}{r_1^3}} \qquad (13)$$

The mass of the Sun is $m_1$= 1.99·10³⁰ kg, $r_1$=6.96·10⁸ m, and $w_1$=1.022·10¹¹ c⁻¹.

The ether linear velocity on the surface is $v(r_1)=w_1 \cdot r_1$= 7.113·10¹⁹ m/s. This velocity is lower than the average speed of amers in ether (6.6·10²¹ m/s [1]) by two orders of magnitude. Thus, the obtained value of the ether wind linear velocity appears to be quite reasonable. For the Earth, $m_1$=5.98·10²⁴ kg, $r_1$=6.38·10⁶ m, and $w_1$=2.001·10¹¹ c⁻¹, v ($r_1$)=1.277·10¹⁸ m/s.

On the basis of vortex gravitation, the value of $w_1$ in any celestial torsion is determined from the condition of the equality of the centrifugal forces and the gravitation forces for a celestial body.

Using Eq.(10), one can calculate the orbits of all satellites, determine the gravity on a surface of any celestial body, and, thereby, the values of gravitational acceleration.

Taking into account the compressibility of ether, e.g. in the isothermal case (T=const), i.e. when

$$c = f(P) = \frac{P}{R \cdot T} \qquad (14)$$

where R is the specific gas constant $R = \dfrac{R_0}{M} = \dfrac{R_0}{m_0 \cdot Na} = 1.972 \cdot 10^{93}$ J·kg$^{-1}$·K$^{-1}$ ($R_0$=8.314 J·mol$^{-1}$·K$^{-1}$ – the absolute gas constant, μ - the ether molecular weight, $m_0$=7·10$^{-117}$ kg – the mass of an amer [1], $N_a$=6.022·10$^{23}$ mol$^{-1}$ – the Avogadro number), after the first equation in system (3) to be solved, we have got a function of the pressure radial distribution. This function, using e.g. the values of $w_1$ and $r_1$ for the Sun, results in a very insignificant change of the density with radius enabling the ether to be considered as an incompressible substance, and thereby, enabling the above-presented formulas to be used.

Let's now find the dependence P(r) solving the first equation of system (3). Taking Eq.(7) into account, we will find that

$$P(r) = P_0 + c \cdot w_1^2 \cdot r_1^3 \cdot \left[ \dfrac{1}{r_1} - \dfrac{1}{r} \right] \quad (15)$$

where $P_0$ is the ether surface pressure. Using the boundary condition $P(\infty) = P_b$, we get $P_0 = P_b - c \cdot w_1^2 \cdot r_1^2$ with $P_b$ being the pressure of free ether.

From the obtained formula for vortex gravitation, it is obvious that in the existing Newton's law of gravitation, instead of the reason of gravity (the gradient of pressure), the consequence of that (i.e. the mass) is used.

### 3. Some Conclusions

In the context of the suggested model of universal vortex gravitation, new principles in modern cosmogony and astrophysics are suggested.

#### 3.1. Black Holes

In 1783, John Mitchell has presented his work, wherein he pointed out that a sufficiently massive and compact star must possess a strong gravitational field, so that it prevents the light from traveling outside. Such objects are called the 'black holes'.

According to the calculations of astrophysicists, the gravitation force capturing the light can be associated with an object having the mass of the Sun and the own radius of 3 km. That is, it is a superdense star being in the state of self-collapse.

On the basis of the theory of vortex ether rotation, the gravity can be defined in any space point. Therefore, it is established by calculations that the black-hole supergravitation is created by the solar torsion inside the Sun in that same distance, i.e. 3 km from its center.

It should be noted that such supergravitation is created by the Earth torsion in a distance of 5 meters from the center.

This suggests that black holes can be of various volumes. Only the black holes with very large own volumes can absorb large celestial bodies.

An exterior observer can fix a black hole only at that moment when the center of this cosmic torsion is not closed by the cosmic substance (which the torsion must absorb since the moment of its origin) yet. After the space substance is concentrated in the torsion center up to the volume which closes the ultrafast zone, this celestial object turns into a usual celestial body – planet, star, etc.

Therefore, the black hole is the center of rotation of the space torsion. As a result of this rotation and the gravitation created, a new celestial body must appear. That is, a black hole is not a collapse of a celestial body, but it is a newly-created space torsion.

### 3.2 Expansion or compression of the Universe?!

At present, the moving of galaxies away from each other is accounted for by the expansion of the Universe. This expansion is thought to start due to the so-called 'Big Bang'.

According to the theory of vortex gravitation and the laws of mechanics, the Universe should be in the state of compression and twisting, not expansion. This is proved under the following conditions:

– The Universe ether and galaxies rotate around the center;

- all the celestial bodies increase their masses permanently.

These regularities are confirmed by astrophysicists: the galaxies rotate around the center of the Universe completing one turnover per **100 billion years [4].** The mass of Earth increases by **$1.6 \cdot 10^{15}$ kg a year [1].**

According to the law of momentum conservation **(m·v = const)**, as the mass of a moving body increases, its velocity decreases, i.e. the orbital velocity of galaxies decreases.

On the other hand, a decrease of the rotation velocity leads to a decrease of the centrifugal force according to the formula:

$$F_c = \frac{m \cdot v^2}{r} \qquad (16)$$

The gravitation force is independent of the orbital velocity, and hence it does not decrease. Thus, the change in the ratio of the two counterbalancing forces occurs in favor of the gravitation force. That is, the galaxies, apart from the orbital motion, have also the radial motion directed toward the center of rotation.

**Consequently, the Universe is compressed and twisted.**

However, the decrease of the distance to the center means the decrease of the radius of the motion orbit, which in turn results in the square increase of the gravitation force (see Eq's (1) or (10)), while the centrifugal forces increase only linearly, Eq.(16). Thus, the closer is the galaxy situated to center of the Universe, the faster it moves toward the center. This accounts for the moving of the galaxies away from each other with the acceleration which equals to the Hubble constant.

It is quite possible that there exists the Universe Black Hole in the central zone of the Universe. Therefore, this zone is invisible.

### 3.3. Densities of the Planets

According to the theory of vortex space rotation, the gravitation force is independent of the masses and densities of the bodies; therefore, the masses of the planets have been determined on the basis of the law of angular momentum conservation, under the following conditions.

The celestial bodies were created in the centers of space torsions by means of the matter accumulation. During this process, the own mass of the object has been increased from an initial value (which is equal to the mass of ether) up to the final one, which is equal to the today's mass of the object. Similarly, the rotation speed of the celestial body has decreased from the initial one (ether rotation) down to the final value, i.e. the today's velocity of rotation of this celestial object.

The calculated densities of the planets are presented in Table 2 in comparison with the commonly accepted data.

The same regularity in the velocity decrease should also take place for the orbital motion of celestial bodies. It should be stressed that the rate of decrease of all the velocities was maximum in the initial period of the creation of planets and stars, because the change of the mass and velocity, according to the law of momentum conservation, is taken into account as a relative one, not absolute. According to the calculations, during the first year of the existence, the mass of the space torsion increased (and, correspondingly, the velocity decreased) by a factor of more than one trillion. During the second year, the velocity of rotation decreased approximately by a factor of two. At present, this slowing-down does not exceed $\sim 10^{-10}$ per year.

The velocity of torsion motion in radial direction, in turn, is also changed directly proportional to the change of the orbital velocity.

Because the space torsion (in the initial period of the existence) must be in a gravitational state corresponding to that of a black hole, the motion of this black hole distinguishes significantly from the motion of large celestial objects. The point is that, in this period, the change of the velocity – both in orbital, and in radial direction – is the maximum. Therefore, the approach of the black holes to the celestial bodies, including their mutual absorption, is possible.

### 3.4 Age of Planets and Sun

The modern theories of interior structure of celestial bodies, as well as planetary cosmogony, use some experimental results as a basis for evaluations of the age of celestial bodies. Among these results, one can mention the investigation of rock age, solar neutrino, and some other data obtained from the studies of the outer layers of the celestial bodies.

Provided that the celestial bodies have been created by means of the cosmic matter accumulation, one can conclude that each inner layer should have its own age, which exceeds the age of the outer layer of the planet or star. Therefore, it is impossible to evaluate the age of the celestial-body interiors from the data concerning the inner rocks or the radiation emitted by these rocks.

In the model of vortex cosmogonic development, the gravitation force and the degree of the planetary matter accumulation are determined only by the rotation velocity of the Earth ether torsion. Neglecting the vortex damping, the values of rotation velocity, gravitation, and mass increase can be considered as constants during the period of the Earth existence.

Accordingly, the age of the planets is determined by the ratio of the planet mass to the corresponding mass increase.

The results of calculations are presented in Table 3.

The ages of the Sun and planets were calculated using the numerical quantities of their densities corresponded to the classical data; the obtained values are identical to each other – **3.75 billion years.**

### 4. Evidences of Vortex Rotation and Gravitation

**The first evidence for the vortex cosmic rotation consists in a generally-known regularity, viz.:**

- The faster the planet rotates around the axis, the greater is its mass and more satellites it possesses.

This regularity convincingly supports the vortex nature of gravitation, because it has the following cause-effect relation:

- The faster the planet rotates, the faster the corresponding ether torsion rotates too. The faster the torsion rotates, the stronger is the force of vortex gravitation. The stronger is the force, the higher is the degree of absorption of cosmic matter by this torsion, and hence the larger is the mass of the created celestial body and the number of its satellites.

Table 1 presents the priority of the planets in their own parameters – rotation velocity, physical volume, and the number of satellites.

The velocity of motion of the Sun surface is higher than that of the planets' surfaces by an order of magnitude.

**The second evidence for the ether vortex rotation consists in the orbital motion of the planets.**

It is known that the velocities of circulation of the planets around the Sun increase inversely with the square of the distance to the rotation center.

Such distribution of the orbital velocities in a uniform continuum occurs only at torsion rotation of the continuum (ether). In the other physical systems, such regularity in the motion velocity distribution for the subjects of one system has not been found so far.

Therefore, we have reached the following conclusion:

- Because the orbital rotation of the solar-system planets corresponds to the torsion rotation of a continuum, the motion of these planets has been caused by rotation of this continuum, i.e. ether. Accordingly, the ether is in the state of torsion rotation.

Newton was apparently the first who realized this regularity in the planet velocities, discovered after the invention of a telescope. He understood that the rotation speed was linked to the gravitation force. Based on this idea, he developed the well-known law of gravity, putting in the formula the not less well-known square of distance to the center of rotation.

**Besides, the obvious fact – viz, that all the celestial bodies, as well as the systems of those, rotate in the Universe permanently – confirms the torsion principle of the world matter existence.**

It should be noted in conclusion that the theory of vortex gravitation will allow it to either define more exactly or even change the solutions of numerous cosmology problems, such as: determination of the gravity at other celestial bodies and the speed of gravitation; the origin of magnetic fields of the planets and stars; the appearance of cosmic torsions; understanding of the approach of galaxies and black holes; explanation of the Zeliger paradox; development of the theory of evolution of the Universe and biological life, both in the past, and in the future; and many other things.

**Acknowledgments.** The author thanks A. A. Velichko for the help in carrying out the mathematical calculations in Section 2.1.

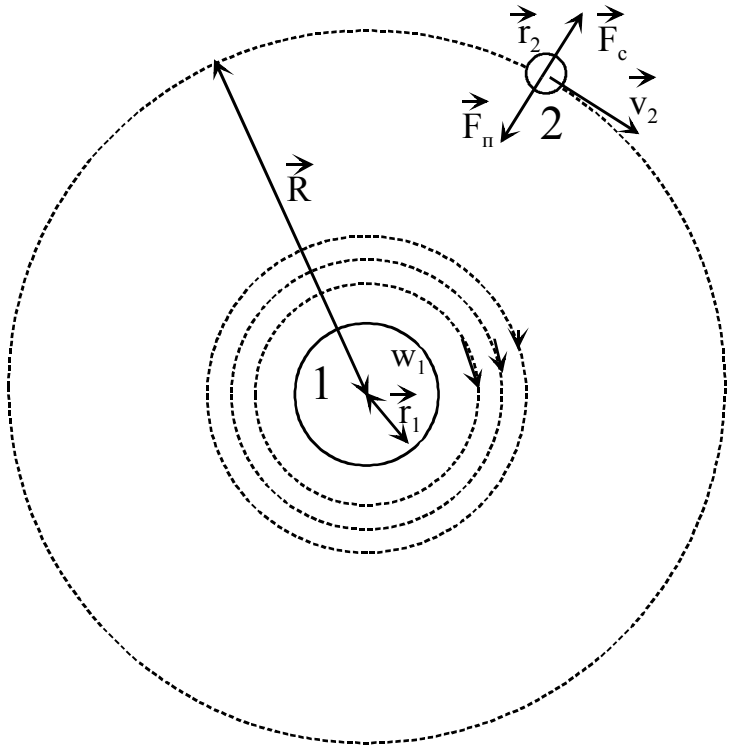

Fig. 1

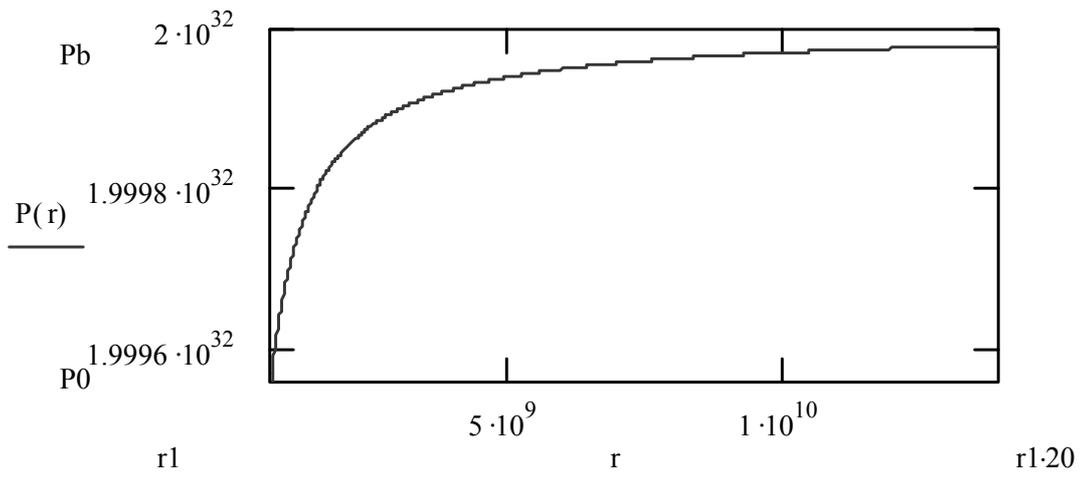

Fig. 2

Fig.1. Two-dimensional model of gravitational interaction of two bodies. The forces are shown acting on body 2: $F_c$ – centrifugal force, $F_п$ – the force of attraction of body 2 from body 1; $v_2$ – linear velocity of body 2 at the orbit, R – the radius of the orbit, $r_1$ – the radius of body 1, $r_2$ – the radius of body 2, $w_1$ – angular velocity of ether rotation on the surface of body 1.

Fig.2. Radial distribution of the ether pressure for the Sun.

Table 1.

|  | Jupiter | Saturn | Uranium | Neptune | Earth | Mars | Pluto | Venus | Mercury |
|---|---|---|---|---|---|---|---|---|---|
| **Surface velocity, V(r)** | 1 | 2 | 3 | 4 | 5 | 6 | 7 | 8 | 9 |
| **Volume** | 1 | 2 | 3 | 4 | 5 | 7 | 9 | 6 | 8 |
| **Number of satellites** | 1 | 2 | 3 | 4 | 6 | 5 | no | no | no |

Table 2.

| Density, kg/m$^3$ | Sun | Earth | Mars | Jupiter | Saturn | Uranium | Neptune |
|---|---|---|---|---|---|---|---|
| **Density in catalogues** | 1400 | 5500 | 4000 | 1300 | 700 | 1500 | 1700 |
| **Density from calculations** | 31000 | 23000 | 20566 | 6000 | 3840 | 5500 | 1000 |

Table 3.

|  | Sun | Earth | Mars | Jupiter | Saturn | Uranium | Neptune |
|---|---|---|---|---|---|---|---|
| **Age, billion year** | 87 | 16 | 18 | 11 | 21 | 18 | 1.6 |